\documentclass[aps,prd,twocolumn,showpacs]{revtex4}

\usepackage{amssymb}

\newcommand{\h}[1]{{}^{\mbox{\,\tiny $\{#1\}\!$}}h}
\newcommand{\pert}[2]{{}^{\mbox{\,\tiny $\{#1\}\!$}}{#2}}
\newcommand{\flow}[2]{{}^{\mbox{\,\tiny $\{#1\}\!$}}{#2}}

\newcommand{\calH}[7]{{{}^{\mbox{\tiny $({#1})\!$}}
{\mathcal H}_{#2}^{#3}{}_{#4}^{#5}{}_{#6}^{#7}}}
\newcommand{\calHtilde}[7]{{{}^{\mbox{\tiny $({#1})\!$}}{\tilde{\mathcal
H}}_{#2}^{#3}{}_{#4}^{#5}{}_{#6}^{#7}}}
\newcommand{\E}[7]{{E_{#1}^{#4}{}_{#2}^{#5}{}_{#3}^{#6}
{}^{}_{#7}}}
\newcommand{\C}[6]{{C_{#1}^{#2}{}_{#3}^{#4}{}_{#5}^{#6}}}

\begin{document}

\title{High-order gauge-invariant
perturbations of a spherical spacetime}

\author{David Brizuela}
\author{Jos\'e M. Mart\'{\i}n-Garc\'{\i}a}
\author{Guillermo A. Mena Marug\'an}
\affiliation{Instituto de Estructura de la Materia,
CSIC, Serrano 121-123, 28006 Madrid, Spain}

\date{\today}

\pacs{04.25.Nx, 04.20.Cv, 95.30.Sf}

\begin{abstract}
We complete the formulation of a general framework for
the analysis of high-order nonspherical perturbations
of a four-dimensional spherical spacetime by including
a gauge-invariant description of the perturbations. We
present a general algorithm to construct these
invariants and provide explicit formulas for the case
of second-order metric perturbations. We show that the
well-known problem of lack of invariance for the
first-order perturbations with $l=0,1$ propagates to
increasing values of $l$ for perturbations of higher
order, owing to mode coupling. We also discuss in
which circumstances it is possible to construct the
invariants.
\end{abstract}

\maketitle

\section{Introduction}
The highly nonlinear character of the Einstein
equations of General Relativity requires the use of
numerical and perturbative techniques to study
problems of astrophysical interest. Processes like the
merge of binary systems (black holes and/or neutron
stars) are promising candidates for gravitational wave
detection in the next decade, and need careful and
detailed simulation to accurately predict the
waveforms that will be observed.

The use of perturbation theory can be important both
as a tool for simulating the evolution of a system in
a state close to a known solution, or as a tool to
extract the information of the amount and form of the
gravitational radiation emitted. First-order
perturbation theory is normally used in both cases,
but it is possible to improve the accuracy of the
results by using perturbation theory to higher orders.

We have recently introduced a general framework to
study generic high-order nonspherical perturbations of
an arbitrary spherical spacetime \cite{BMM06}. It can
be considered as a generalization of the formalism of
Gerlach and Sengupta (GS) for first-order
perturbations around spherical symmetry \cite{GeSe79},
in the sense that we use the same concepts and
techniques: the background spacetime is decomposed as
the product of a 2-dimensional Lorentzian manifold and
the 2-sphere; covariant notations are used on both
submanifolds; and all perturbations are covariantly
decomposed as series in tensor harmonics. However, the
GS formalism also uses a gauge-invariant description,
whereas our presentation in Ref. \cite{BMM06} employed
a fixed gauge (the Regge-Wheeler gauge \cite{ReWh57}).
The present article fills in this gap and completes
the construction of the high-order generalization of
the GS formalism by showing how one can construct
gauge-invariant high-order perturbations. This allows
one to treat the perturbations in any gauge while
dealing with quantities and expressions that present
the same form for all physically equivalent
perturbations.

The issue of perturbative gauge invariance is closely
related, though not equal, to the coordinate
independence of General Relativity and has been dealt
with in different ways already in first-order
perturbation theory. The pioneering works by Sachs
\cite{Sac64}, Moncrief \cite{Mon73}, and Stewart and
Walker \cite{StWa74} established most of the basic
ideas of the subject at first order, but offered
different points of view about gauge invariance, what
sometimes obscures the interpretation of the results
and complicates any attempt of generalizing the
analysis to higher orders. Bruni and collaborators
have shown \cite{BMM97} that the geometrical approach
of Sachs \cite{Sac64}, already rather constraining at
first order, becomes even more restrictive at higher
orders and hence only highly symmetric scenarios can
be described in this approach. Here we will show that
the alternative approach by Moncrief \cite{Mon73}
allows a greater and more convenient flexibility, so
that we will adhere to it in order to construct metric
gauge invariants at both first and second order around
a spherical spacetime.

This article, which can thus be considered a
continuation of Ref. \cite{BMM06} (referred to as
Paper I in the following), is organized as follows.
Sec. \ref{section::gauge} reviews the concept of gauge
transformations and Sec. \ref{section::invariance}
discusses the different notions of gauge invariance
that have been employed in the literature, introducing
a general method for the construction of gauge
invariants. Sec. \ref{section::spherical} describes
the notation adopted for spherically symmetric
spacetimes and presents a brief summary on the
structure of tensor harmonics (see Paper I for a more
detailed description). Sec. \ref{section::sphericalGI}
constructs gauge-invariant nonspherical perturbations
around a spherical spacetime, using the method
introduced in Sec. \ref{section::invariance}, and is
the central section of the paper. Finally, Sec.
\ref{conclusions} contains our conclusions. Appendix
\ref{coupling} is devoted to the study of mode
coupling and discusses in which cases the high-order
gauge invariants can be consistently constructed. A
particular case, already studied in the literature
\cite{GaPr00}, is described in Appendix \ref{GaPr}.

\section{Gauge freedom in perturbation theory}
\label{section::gauge}

Perturbation theory in General Relativity considers a
family of spacetimes $(M_\varepsilon, g_\varepsilon)$
in which associated families of tensor fields
$\Omega_\varepsilon$ are defined. The $\varepsilon=0$
members of those families are referred to as the {\it
background} members and will be denoted without a
subscript, e.g. the background metric is $g\equiv
g_0$. All manifolds $M_\varepsilon$ are assumed to be
diffeomorphic. The main issue in perturbation theory
is comparing tensor fields for a given nonzero value
of $\varepsilon$ with their background counterparts.
There exists diffeomorphism invariance on each of the
manifolds $M_\varepsilon$ but, in addition, there is
no preferred point-to-point identification mapping
between any two such manifolds, so that the comparison
of two tensor fields with different values of
$\varepsilon$ is not an invariantly defined concept.
This is the origin of the so-called {\it gauge
freedom} in perturbation theory \cite{StWa74}.

Let us call {\it a gauge} $\phi_\varepsilon$ a family
of point-to-point identification diffeomorphisms from
the background manifold to $M_\varepsilon$:
\begin{equation}
\phi_\varepsilon : \quad M \quad \longrightarrow \quad
M_\varepsilon .
\end{equation}
Given a gauge $\phi_\varepsilon$ we can now pull-back
a generic tensor $\Omega_\varepsilon$ on
$M_\varepsilon$ to a tensor
$\phi^*_\varepsilon\Omega_\varepsilon$ on $M$. This
latter tensor can be compared with the background
member $\Omega$ (at each point $x\in M$), resulting in
a $\phi$-dependent concept of what a perturbation
means. Assuming smooth dependence of all structures in
$\varepsilon$, we can define the perturbative
expansion
\begin{equation}\label{expansion}
\phi_\varepsilon^*\Omega_\varepsilon \equiv \Omega +
\sum_{n=1}^{\infty}\frac{\varepsilon^n}{n!}
\Delta^n_\phi[\Omega],
\end{equation}
where all terms of the equation are defined at the
same point of the background manifold $M$. In
particular the perturbations $\Delta_\phi^n[\Omega]$
are tensor fields on $M$. The notation $\Delta_\phi$
stresses the fact that, in general, it is not possible
to define a perturbation without explicitly indicating
which gauge $\phi$ is used. For instance, the
statement that a perturbation vanishes is generically
meaningless unless one specifies the gauge in which
this occurs. Note also that the infinite series in Eq.
(\ref{expansion}) arises from the simultaneous
dependence on $\varepsilon$ of both
$\Omega_\varepsilon$ and $\phi_\varepsilon$.

One then has to face the question of how the
perturbations $\Delta_\phi^n[\Omega]$ vary under a
change of gauge from $\phi_\varepsilon$ to, let's say,
$\psi_\varepsilon$ while keeping unaltered the family
of tensors $\Omega_\varepsilon$. Such a {\it gauge
transformation} will be described by a family
$\chi_\varepsilon$ of diffeomorphisms on the
background manifold
\begin{equation}\label{gaugetransf}
\chi_\varepsilon \equiv \phi_\varepsilon^{-1} \circ
\psi_\varepsilon : \quad M \quad \longrightarrow \quad
M ,
\end{equation}
which clearly satisfy
\begin{equation}\label{transformation}
\psi^*_\varepsilon \Omega_\varepsilon =
\chi_\varepsilon^* \, \phi^*_\varepsilon
\Omega_\varepsilon .
\end{equation}
We emphasize that $\chi_\epsilon$ is not a gauge, but
a gauge transformation.

The theory of expansions of families of general
diffeomorphisms on a given manifold has been developed
in Ref. \cite{BMM97}. The most important result is
that any one-parameter family of diffeomorphisms, in
particular the gauge transformation
$\chi_\varepsilon$, is equivalent to an infinite set
of flows (a flow is a one-parameter group of
diffeomorphisms) $\{\Phi^{(1)}_\varepsilon,...,
\Phi^{(m)}_{\varepsilon^m/m!},...\}$, so that
$\chi_\varepsilon$ can be expressed in the following
way
\begin{equation}\label{knight}
\chi_\varepsilon=...
\circ\flow{m}{\Phi}_{\varepsilon^m/m!}\circ...
\circ\flow{2}{\Phi}_{\varepsilon^2/2}
\circ\flow{1}{\Phi}_{\varepsilon}.
\end{equation}
From the definition of the Lie derivative, it is
straightforward to see that the pull-back of any flow
$\Phi_\varepsilon$ acting on a generic tensor $T$ can
be expanded as
\begin{equation}\label{groupexpand}
\Phi_\varepsilon^*T=\sum_{n=0}^{\infty}
\frac{\varepsilon^n}{n!}{\mathcal L}_\xi^nT,
\end{equation}
where the vector $\xi$ is the generator of the flow.
Making use of this formula and the decomposition
(\ref{knight}), we can expand the right-hand side of
Eq. (\ref{transformation}) in a power series, so that
two gauge choices are related by
\begin{eqnarray}\label{Psiexpansion}
\psi^*_\varepsilon \Omega_\varepsilon&=&
\sum_{k_1=0}^{\infty}\sum_{k_2=0}^{\infty}...
\sum_{k_m=0}^{\infty}...
\frac{\varepsilon^{k_1+2k_2+...+mk_m+...}}
{2^{k_2}...(m!)^{k_m}...k_1!k_2!...k_m!...}
\nonumber\\
&\times& {\mathcal L}^{k_1}_{\flow{1}{\xi}} {\mathcal
L}^{k_2}_{\flow{2}{\xi}}... {\mathcal
L}^{k_m}_{\flow{m}{\xi}}... \phi^*_\varepsilon
\Omega_\varepsilon.
\end{eqnarray}
The vector fields $\{\flow{1}{\xi},\flow{2}{\xi},...,
\flow{m}{\xi},...\}$ are the corresponding generators
of the flows $\{\flow{1}{\Phi},
\flow{2}{\Phi},...,\flow{m}{\Phi},...\}$. Substituting
expansion (\ref{expansion}) in the above equation, we
get \cite{SoBr98}
\begin{eqnarray}\label{ngaugetrans}\nonumber
\Delta_\psi^n[\Omega] - \Delta_\phi^n[\Omega]\!\!
&=&\!\!\!\sum_{m=1}^{n}\frac{n!}{(n-m)!}\!\!\sum_{(K_m)}
\!\frac{1}{2^{k_2}...(m!)^{k_m}k_1!...k_m!}\\
&\times&\quad{\mathcal L}_{\flow{1}{\xi}}^{k_1}...
{\mathcal L}_{\flow{m}{\xi}}^{k_m}
\Delta^{n-m}_\phi[\Omega],
\end{eqnarray}
where we have defined \[ (K_m)=\left\{(k_1,...,k_m)
\in\mathbb{N}^m; \quad \sum_{i=1}^{m}i
k_i=m\right\}.\] Up to third order we obtain
\begin{eqnarray}
\Delta_\psi[\Omega]-\Delta_\phi[\Omega]&=&
{\mathcal L}_{\flow{1}{\xi}}\Omega,\\
\Delta^2_\psi[\Omega]-\Delta^2_\phi[\Omega]&=&
\left({\mathcal L}_{\flow{2}{\xi}}+{\mathcal
L}^2_{\flow{1}{\xi}}\right)\Omega\nonumber\\
&+&2{\mathcal L}_{\flow{1}{\xi}}\Delta_\phi[\Omega],\\
\Delta^3_\psi[\Omega]-\Delta^3_\phi[\Omega]&=&
\left({\mathcal L}_{\flow{3}{\xi}} +{\mathcal
L}^3_{\flow{1}{\xi}}+3 {\mathcal
L}_{\flow{1}{\xi}}{\mathcal L}_{\flow{2}
{\xi}}\right)\Omega\nonumber\\ &+&3\left({\mathcal
L}_{\flow{2}{\xi}}+{\mathcal
L}^2_{\flow{1}{\xi}}\right)\Delta_\phi[\Omega]\nonumber
\\
&+&3{\mathcal L}_{\flow{1}{\xi}}\Delta^2_\phi[\Omega].
\end{eqnarray}
These formulas describe the effect of general gauge
transformations on any high-order perturbation of a
generic background tensor $\Omega$. They contain all
the information needed to analyze the issue of gauge
transformations in perturbation theory.

In practical applications we can keep terms only up to
a finite order $\varepsilon^n$, which projects the
full group ${\cal G}$ of gauge transformations into a
truncated group $\pert{n}{\cal G}$ of $n$th-order
gauge transformations. Each of these transformations
is described by a collection of $n$ vector fields
$\{\flow{1}{\xi}, ..., \flow{n}{\xi}\}$, and we will
say that it is {\em pure} $m$th-order if all those
vectors are zero except for $\flow{m}{\xi}$. The
important point is that, in general, composition of
pure $m$th-order transformations is not pure
$m$th-order, unless $m=n$. For example, the
composition of two generic second-order
transformations described by
$\{\flow{1}{\bar\xi},\flow{2}{\bar\xi}\}$ and
$\{\flow{1}{\hat\xi},\flow{2}{\hat\xi}\}$ is described
by the pair
\begin{equation} \label {coupling1}
\{ \flow{1}{\bar\xi} + \flow{1}{\hat\xi} , \quad
   \flow{2}{\bar\xi} + \flow{2}{\hat\xi} +
   [ \flow{1}{\bar\xi} , \flow{1}{\hat\xi} ]
\} ,
\end{equation}
and hence the subset of pure first-order
transformations $\{\flow{1}{\xi},0\}$ is not a
subgroup of $\pert{2}{\cal G}$. In fact, the group
$\pert{1}{\cal G}$ is not equivalent to this subset,
but only to a truncated form of it, and therefore it
is important to distinguish between first-order
transformations $\{\flow{1}{\xi}\}$ and pure
first-order transformations
$\{\flow{1}{\xi},0,...0\}$. In general, the set
$\pert{n}{\cal G}$ of all $n$th-order gauge
transformations is a group, but the subset of all pure
$m$th-order transformations is not. The only exception
is the reduced case $m=n$ of transformations of the
form $\{0,...,0,\flow{n}{\xi}\}$, in which only a
single linear term in Eq. (\ref{ngaugetrans}) survives
(this includes first-order perturbation theory as the
case $m=n=1$). There are more general subgroups of a
given $\pert{n}{\cal G}$, like the subgroup of
transformations of the form $\{0,...,0,\flow{m}{\xi},
\flow{m+1}{\xi}, ..., \flow{n}{\xi}\}$, but they have
less interest for our discussion and will not be
considered in this work.

\section{The notion of gauge invariance}
\label{section::invariance}

Once we have defined the concept of gauge
transformation in Eq. (\ref{gaugetransf}), we discuss
now the associated notion of {\em gauge invariance} of
a family of tensors $\Omega_\varepsilon$ under a group
of gauge transformations. We will then find the
inherited gauge invariance of the perturbations
$\Delta^n_\phi[\Omega]$ under the respective truncated
version of that group.

The most natural definition of gauge invariance was
given by Sachs \cite{Sac64}: A tensor family
$\Omega_\varepsilon$ is {\em identification gauge
invariant} (IGI) if the pull-back of its members to
the background manifold is independent of the gauge,
though the result still depends on $\varepsilon$. That
is, $\phi_\varepsilon^*\Omega_\varepsilon =
 \psi_\varepsilon^*\Omega_\varepsilon$ for all
gauges $\phi_\varepsilon, \psi_\varepsilon$. This can
also be interpreted as the invariance under the full
group ${\cal G}$ of gauge transformations.
Perturbatively, a tensor family $\Omega_\varepsilon$
is IGI up to order $n$ if and only if
$\Delta^m_\phi[\Omega]=\Delta^m_\psi[\Omega]$ for all
$m\leq n$ and all gauges $\phi$, $\psi$ \cite{BMM97}.
Again, this is equivalent to the requirement of
invariance under the truncated group $\pert{n}{\cal
G}$ defined above. This definition turns out to be too
restrictive because, as it is well known
\cite{Sac64,StWa74}, only perturbations of vanishing
tensors, constant scalars, or constant linear
combinations of products of delta tensors can be IGI
in first-order perturbation theory, since these are
the only tensors with zero Lie derivative along every
vector field. For higher, $n$th-order perturbations
the problem becomes even worse because, apart from the
background quantities, all of the $m$th-order
perturbations with $m<n$ must also be of the form that
we have commented \cite{BMM97}. In principle, this
restricts to a very narrow physical scenario the
possibilities that are left.

Other forms of gauge invariance can be defined using
subgroups of ${\cal G}$ or $\pert{n}{\cal G}$. For
example invariance with respect to the reduced
subgroup of pure $n$th-order transformations has been
used in the past \cite{CPM80}, by fixing the gauge
perturbatively at all orders $1,...,n-1$, but not at
order $n$.

Only perturbations of highly symmetric backgrounds
admit a complete description in terms of IGI
variables. Even at first order, significant
limitations have been found: Stewart and Walker showed
that, for vacuum spacetimes, only backgrounds with
Petrov type D are possible \cite{StWa74}, which
fortunately includes the Kerr spacetime. In cosmology,
only perturbations of static Friedman-Robertson-Walker
(FRW) backgrounds can be described in terms of IGI
variables \cite{Ste90}. For spherical backgrounds with
matter, only first-order perturbations with axial
polarity admit such a description \cite{Car00,Nol04},
but not the complementary set of polar perturbations.
This latter result is specially relevant for us,
because we want to construct high-order gauge
invariants of a spherical spacetime and in Paper I we
saw that the polarities mix already at second order:
hence there is no hope of getting a purely IGI
description in general. Note that vacuum \cite{Mon73}
and electro-vacuum \cite{Mon75} spacetimes with
spherical symmetry are very special cases (in
particular included in the cited result for type D
spacetimes), for which the programme of construction
of gauge invariants can be further developed
\cite{SarbachTiglio}.

On the other hand, when describing gravitational
radiation in a vacuum, the Weyl tensor provides all
the relevant geometrical information, and therefore
many investigations employ it as the basic object to
be perturbed. Furthermore, the Weyl tensor defines a
set of principal null directions, so that it becomes
natural to decompose it using the Newman-Penrose
formalism. The analysis of IGI is then simplified, but
an additional type of gauge invariance is introduced,
called {\em tetrad gauge invariance}, which requires
invariance under (the 6-parametric Lorentz group of)
transformations among null tetrads \cite{StWa74,
CaLo99, Nol04}. We will not use this approach in this
paper, but instead appeal to a different and more
general notion of gauge invariance in which one makes
use of an additional geometrical structure: a
privileged gauge $\overline\phi_\varepsilon$. The
basic idea is that, given a family of tensors
$\Omega_\varepsilon$, one can select a privileged
gauge to extract the physical information contained in
this family and express this information in terms of
the pull-back of $\Omega_\varepsilon$ in an arbitrary
gauge. In other words, the gauge invariant is defined
as the function(al) that provides the value of
$\overline\phi_\varepsilon^*\Omega_\varepsilon$ in a
generic gauge:
\begin{equation}
\overline\phi_\varepsilon^*\Omega_\varepsilon =
F\left[ \phi_\varepsilon^*\Omega_\varepsilon \right] .
\end{equation}
So, the gauge invariant is now supplied by the
function(al) $F$ rather than by the family of tensors
$\Omega_\varepsilon$ itself, as was the case for IGI.

This notion of gauge invariants is similar to that of
the constants of motion defined in Mechanics by the
particular values that the variables of the system
take at some fixed instant of time \cite{Goldstein},
or even to the notion of {\em evolving constants of
motion} recently introduced in Quantum Gravity
\cite{Rovelli} (although in that case one ought to
consider a family of privileged gauges parameterized
by a set of real numbers, rather than just one of them
$\overline\phi_\varepsilon$). In spite of appearing
counterintuitive at first, this notion can be very
useful in those cases in which the computations can be
carried out to completion, i.e. when one can obtain
the explicit expression of the invariants in terms of
gauge-dependent quantities, in our context, or in
terms of time-dependent variables, in Mechanics. In
other words, we need to determine the explicit form of
the gauge transformation
$F=\chi_\varepsilon^*=\phi_\varepsilon^{-1}\circ
\overline\phi_\varepsilon$ for arbitrary
$\phi_\varepsilon$. Whether this is possible or not
essentially depends on the choice of gauge
$\overline\phi_\varepsilon$.

In practice, the privileged gauge is defined by
imposing some conditions $R_\varepsilon$ on the
pull-back
$\overline\phi_\varepsilon^*\tilde\Omega_\varepsilon$
of a particular tensor $\tilde\Omega_\varepsilon$.
Therefore, $\overline\phi_\varepsilon$ will be
characterized as the gauge in which the tensor
$\overline\phi_\varepsilon^*\tilde\Omega_\varepsilon$
satisfies some specific requirements. For this method
to work satisfactorily, this privileged gauge choice
has to be {\em rigid}. This means that the conditions
$R_\varepsilon[\overline\phi_\varepsilon^*
\tilde\Omega_\varepsilon]=0$ must fix uniquely the
gauge $\overline\phi_\varepsilon$, and so any further
gauge transformation will violate those conditions.

In perturbation theory, the invariants will then be
the combinations
$\pert{n}{F}[\{\Delta_\phi^m[\Omega]\}]$ obtained by
performing a gauge transformation from the
perturbations defined on a generic gauge
$\Delta_\phi^m[\Omega]$ to those defined in the rigid
gauge $\Delta_{\bar\phi}^m[\Omega]$. This kind of
combination of perturbations have been characteristic
of this approach to gauge invariance, starting with
the pioneering work of Moncrief \cite{Mon73} for
nonspherical perturbations of Schwarzschild, where the
Regge-Wheeler (RW) gauge was implicitly used as the
privileged gauge. His work was later generalized by GS
\cite{GeSe79} to nonspherical perturbations of any
spherical background, also implicitly using the RW
gauge. The same procedure has been employed by Bardeen
\cite{Bar80}, Stewart \cite{Ste90}, and many other
authors in their study of perturbations of FRW
cosmologies. It can also be found in several recent
investigations of second-order perturbations of vacuum
\cite{CaLo99,GaPr00} or cosmological backgrounds
\cite{MaWa04}.

For instance, the first-order gauge invariants of a
generic tensor $\Omega$ will be given by
\begin{equation}\label{firstorderGI}
F\big[\Delta_\phi[\Omega]\big]\equiv
\Delta_\phi[\Omega]+{\mathcal L}_{p}\Omega,
\end{equation}
where $p^{\mu}$ is the vector field generating the
first-order gauge transformation from $\phi$ to
$\overline\phi$, so that this vector contains now
information about our choice of privileged gauge
$\overline\phi$. Again, in practical applications the
gauge $\overline\phi$ is selected by imposing some
rigid conditions $R$ on the perturbations
$\Delta_{\bar\phi}[\tilde\Omega]$ for some specific
tensor $\tilde\Omega$, and such that no residual
freedom is left in the choice of gauge:
\begin{equation}
R\left[\Delta_{\bar\phi}[\tilde\Omega]\right] = 0.
\end{equation}
In this way we get the equations
\begin{equation}\label{cond}
R\left[\Delta_{\phi}[\tilde\Omega]+{\cal L}_p
\tilde\Omega\right] = 0
\end{equation}
which must be solved for $p^{\mu}$ in terms of
$\Delta_\phi[\tilde\Omega]$. Substituting the vector
$p^{\mu}$ obtained in this way, expressions
(\ref{firstorderGI}) provide gauge invariants by
construction. Note that when $\Omega=\tilde\Omega$,
some of those expressions (or combinations of them)
are trivial identities [equivalent to the requirements
(\ref{cond})]. This method for the determination of
invariants can be straightforwardly generalized to
higher perturbative orders, as we will see in the
following case.

Since metric perturbations play a central role in our
analysis, we choose the background metric $g_{\mu\nu}$
as the tensor $\tilde\Omega$ on which one imposes the
conditions to fix the privileged gauge. We introduce
the following compact notation for the perturbations
of the metric:
\begin{eqnarray}
\pert{n}{h}_{\mu\nu}\equiv\Delta^n_{\phi}
[g_{\mu\nu}], \\
\pert{n}{{\mathcal K}}_{\mu\nu}\equiv
\Delta^n_{\overline\phi}[g_{\mu\nu}],
\end{eqnarray}
for a generic gauge $\phi$ and our privileged one
$\overline\phi$, respectively. At first order we have
that expressions (\ref{firstorderGI}) for the metric
become
\begin{equation}\label{firstordeMI}
{\mathcal K}_{\mu\nu}\equiv h_{\mu\nu} +{\mathcal
L}_{p}g_{\mu\nu}.
\end{equation}
The vector $p^{\mu}$ is determined by demanding some
conditions $R[{\mathcal K}_{\mu\nu}]=0$ which
characterize the gauge $\overline\phi$ at first order.
Then, the vector $p^{\mu}$ is determined in terms of
the components of $h_{\mu\nu}$ by solving the
equations
\begin{equation}\label{1metric}
R[h_{\mu\nu}+{\mathcal L}_{p}g_{\mu\nu}]=0.
\end{equation}
This completes the definition (\ref{firstordeMI}) of
the gauge invariant ${\mathcal K}_{\mu\nu}$ as a
function of $h_{\mu\nu}$.

Nonetheless we note that, owing to the presence of the
Lie derivative, Eqs. (\ref{1metric}) contain
derivatives of the vector $p^{\mu}$, so that their
solution will involve in general integrals of the
metric perturbations. Only when $p^{\mu}$ can be
determined explicitly in an amenable way from the
metric perturbations we will have a useful form of
gauge invariants. This fact will depend on the choice
of the privileged gauge. In particular, we will see
later that around spherical backgrounds the
requirement of getting explicit and nonintegral
expressions for the harmonic components of the vector
$p^{\mu}$ will almost uniquely single out the RW
gauge. We also point out that the same vector
$p^{\mu}$, obtained by solving Eqs. (\ref{1metric}),
can now be employed to define the gauge invariants
associated with any other tensor $\Omega$ as in Eq.
(\ref{firstorderGI}). In addition, note that we can
still interpret ${\mathcal K}_{\mu\nu}$ as (the value
of) the metric perturbations expressed in the rigid
gauge $\overline\phi$ which satisfies conditions
(\ref{1metric}).

At higher orders, and once a rigid gauge is chosen via
some conditions $\pert{m}{R}$ for all $m\leq n$, one
can obtain the $n$th-order metric invariants as
\begin{equation}\label{highorderGI}
\pert{n}{{\mathcal
K}}_{\mu\nu}\equiv\pert{n}{h}_{\mu\nu} +{\mathcal
L}_{\flow{n}{p}}g_{\mu\nu}+ \pert{n}{\mathcal
H}_{\mu\nu}.
\end{equation}
Since this equality reflects the effect of a gauge
transformation, the source $\pert{n}{\mathcal
H}_{\mu\nu}$ is explicitly given by Eq.
(\ref{ngaugetrans}) and depends on lower-order vectors
$\flow{m}{p}^{\mu}$ and perturbations
$\pert{m}{h}_{\mu\nu}$ with $m<n$, but not on
$\flow{n}{p}^{\mu}$. Besides, we remember that the
source vanishes at first order ($\pert{1}{{\mathcal
H}}_{\mu\nu}=0$). On the other hand, the equation that
one has to solve iteratively in order to determine the
gauge vectors $\flow{m}{p}^{\mu}$, from $m=1$ to
$m=n$, takes now the expression
\begin{equation}\label{forpn}
\pert{m}{R}\big[\pert{m}{h}_{\mu\nu} +{\mathcal
L}_{\flow{m}{p}}g_{\mu\nu}+\pert{m}{\mathcal
H}_{\mu\nu}\big]=0.
\end{equation}
In particular, when all the conditions $\pert{m}{R}$
have the same linear functional dependence on their
arguments (for instance because they arise from the
perturbative expansion of just one set of exact linear
gauge conditions on the metric), Eq. (\ref{forpn})
will have the form (\ref{1metric}) but with the source
term $\pert{m}{R}[\pert{m}{\mathcal H}_{\mu\nu}]$.
Therefore, the solutions of these equations will be
constructed essentially in the same way.

Nakamura has suggested a similar approach \cite{Nak03}
to construct high-order gauge invariants. He starts
from the basic assumption that a splitting equivalent
to Eq. (\ref{highorderGI}) is given from the outset,
separating the metric perturbation
$\pert{n}{h}_{\mu\nu}$ into its gauge-invariant part
$\pert{n}{\mathcal K}_{\mu\nu}$ and gauge-variant part
(containing the vectors $\flow{m}{p}^{\mu}$), with the
vectors $\flow{m}{p}^{\mu}$ satisfying some set of
requirements. No proposal is made, however, on how
such a splitting can be attained. Our scheme goes
beyond that proposal, giving a constructive and
general prescription to generate the vectors
$\flow{m}{p}^{\mu}$ from the choice of a rigid gauge,
in such a way that the requirements imposed on
$\flow{m}{p}^{\mu}$ are automatically fulfilled.

After determining the vectors
$\{\flow{1}{p}^{\mu}(h),...,\flow{n}{p}^{\mu}(h)\}$,
the perturbations of any tensor field, and in
particular those of the stress-energy tensor
$\pert{n}{\Psi}_{\mu\nu}$, can be taken to its
gauge-invariant form $\pert{n}{T}_{\mu\nu}$ just by
applying a gauge transformation parameterized by the
above vectors:
\begin{eqnarray}\nonumber
\pert{n}{T}_{\mu\nu}&=&\sum_{m=0}^{n}
\frac{n!}{(n-m)!}\sum_{(K_m)}\frac{1}{2^{k_2}...(m!)^{k_m}
k_1!...k_m!}\\
&\times&\quad{\mathcal
L}_{\flow{1}{p}}^{k_1}...{\mathcal
L}_{\flow{m}{p}}^{k_m}\pert{n-m}{\Psi}_{\mu\nu}.
\end{eqnarray}
In this way we will get a tensor
$\pert{n}{T}_{\mu\nu}(\Psi,h)$ whose dependence on the
perturbations $\pert{m}{\Psi}_{\mu\nu}$ and
$\pert{m}{h}_{\mu\nu}$ ($m\leq n$) will not change
when any gauge transformation is applied to them. In
Sec. \ref{section::sphericalGI} we will use these
techniques to compute the metric and matter gauge
invariants for perturbations of a spherical background
spacetime.

\section{Spherically symmetric spacetimes}
\label{section::spherical}

In the following we will restrict our discussion to
spherically symmetric backgrounds and employ the
notation introduced by GS \cite{GeSe79}. We decompose
the background manifold $\cal M$ as the product ${\cal
M}^2\times S^2$, where ${\cal M}^2$ is a
two-dimensional Lorentzian manifold and $S^2$ is the
two-sphere. Without loss of generality, any
spherically symmetric metric and stress-energy tensor
can be written in the block-diagonal form
\cite{GeSe79}:
\begin{eqnarray} \label{sphericalgdecomposition}
g_{\mu\nu}(x^D, x^d)dx^\mu dx^\nu &=& g_{AB}(x^D)dx^A
dx^B
\nonumber\\
&+& r^2(x^D)\,
\gamma_{ab}(x^d)dx^a dx^b ,  \\
\label{sphericaltdecomposition} T_{\mu\nu}(x^D,
x^d)dx^\mu dx^\nu &=& t_{AB}(x^D)dx^A dx^B
+\frac{1}{2} r^2(x^D) \nonumber\\
&\times& Q(x^D)\,\gamma_{ab}(x^d)dx^a dx^b,
\end{eqnarray}
where Greek indices, capital Latin indices, and
lowercase Latin indices take values in the ranges
$\{0, 1, 2, 3\}$, $\{0, 1\}$, and $\{2, 3\}$,
respectively. In this way, $g_{AB}$ is the metric of
the manifold ${\cal M}^2$, whereas $\gamma_{ab}$ is
the round metric on the sphere, and $r$ is a scalar
function defined on ${\cal M}^2$. In order to avoid
working explicitly with logarithms of the function
$r$, it is usual to define
\begin{equation}
v_A\equiv \frac{r_{,A}}{r}=(\ln r)_{,A}.
\end{equation}
We will use the following notation for the covariant
derivatives associated with the different introduced
metrics:
\begin{equation}
g_{\mu\nu;\rho}=0,\qquad
g_{AB|D}=0,\qquad\gamma_{ab:d}=0.
\end{equation}

In Paper I we generalized the so-called
Regge-Wheeler-Zerilli basis of harmonics
\cite{ReWh57,Zer70*} to any rank. Specifically, we
introduced the following two independent rank $s$
symmetric and traceless tensors on the sphere
\begin{eqnarray}
Z_l^m{}_{a_1...a_s}&\equiv&(Y_l^m{}_{:a_1...a_s})^{\rm
STF},
\\
X_l^m{}_{a_1...a_s}&\equiv&
\epsilon_{(a_1}{}^bZ_l^m{}_{ba_2...a_s)},
\end{eqnarray}
where $Y_l^m$ are the scalar harmonics and the
superscript STF means the symmetric and tracefree
part. This definition is valid for $-l\leq m\leq l$
and $1\leq s\leq l$. In all other cases, these
harmonics are defined to be identically zero, except
for $s=0$, when $Z_l^m\equiv Y_l^m$. These tensor
harmonics provide two families with distinct polarity:
$Z_l^m{}_{a_1...a_s}$ is polar, whereas
$X_l^m{}_{a_1...a_s}$ is axial. Under a parity
transformation, the polar and axial harmonics change
sign as $(-1)^l$ and $(-1)^{l+1}$, respectively.

A basis for tensors of rank $s$ on the sphere is then
formed by the two tensors $\{Z_l^m{}_{a_1...a_s},
X_l^m{}_{a_1...a_s}\}$ and some linear combinations of
products between a basis of tensors of rank $(s-2)$,
on the one hand, and the metric $\gamma_{ab}$ and the
Levi-Civita tensor on the sphere $\epsilon_{ab}$, on
the other hand.

\section{Gauge invariant perturbations of a spherical
background} \label{section::sphericalGI}

\subsection{Harmonic decompositions}

When working on spherical backgrounds, it is
convenient to decompose our geometrical objects in
series of spherical harmonics in order to remove all
angle dependencies from the equations of the problem.
We therefore adopt from now on the following
decomposition for the different tensors of interest:
\begin{widetext}
\begin{eqnarray}\label{metricdec}
\h{n}_{\mu\nu} &\equiv& \sum_{l,m} \left(
\begin{array}{cc}
\pert{n}{H}{}_l^m\!{}_{AB} \; Z_l^m &
\pert{n}{H}{}_l^m\!{}_A \; Z_l^m{}_b
+\pert{n}{h}{}_l^m\!{}_A \; X_l^m{}_b \\
{\rm Sym.}& \;\pert{n}{K}_l^m \; r^2\gamma_{ab} Z_l^m
+ \pert{n}{G}_l^m \; r^2Z_l^m\!{}_{ab} +
\pert{n}{h}_l^m \; X_l^m\!{}_{ab}
\end{array}
\right), \\
\label{metricinvariants} \pert{n}{{\mathcal
K}}_{\mu\nu} &\equiv& \sum_{l,m} \left(
\begin{array}{cc}
\pert{n}{{\mathcal K}}_l^m{}_{AB} \; Z_l^m &
\pert{n}{\cal K}{}_l^m\!{}_A \; Z_l^m{}_b +
\pert{n}{\kappa}_l^m{}_A \; X_l^m{}_b \\
{\rm Sym.}& \;\pert{n}{{\mathcal K}}_l^m \;
r^2\gamma_{ab} Z_l^m + \pert{n}{\tilde{\cal K}}_l^m \;
Z_l^m\!{}_{ab} + \pert{n}{{\kappa}}_l^m \;
X_l^m\!{}_{ab}
\end{array}
\right), \\
\label{gaugedec} \pert{n}{\cal H}_{\mu\nu} &\equiv&
\sum_{l,m} \left(
\begin{array}{cc}
\pert{n}{\cal H}{}_l^m\!{}_{AB} \; Z_l^m &
\pert{n}{\cal H}{}_l^m\!{}_A \; Z_l^m{}_b
+\pert{n}{\check{h}}{}_l^m\!{}_A \; X_l^m{}_b \\
{\rm Sym.}& \; \pert{n}{{\cal H}}_l^m \;
r^2\gamma_{ab} Z_l^m + \pert{n}{\tilde{\cal H}}_l^m \;
Z_l^m\!{}_{ab} + \pert{n}{\check{h}}_l^m \;
X_l^m\!{}_{ab}
\end{array}
\right), \\ \label{gaugeP} \pert{n}{p}_\mu dx^\mu
&\equiv & \sum_{l,m}
\left[\pert{n}{P}_l^m\!{}_A\,Z_l^m\,
dx^A+r^2\left(\pert{n}{P}_l^m\,Z_l^m{}_a+
\pert{n}{q}_l^m\,X_l^m{}_a\right)dx^a\right].
\end{eqnarray}
\end{widetext}
The notation ``Sym.'' indicates that the considered
matrices are symmetric. In the most general case, the
label $l$ runs over all nonnegative integers, whereas
$m$ has the usual restriction $-l\leq m\leq l$. Note
also that we have adopted the convention that polar
(axial) harmonic coefficients are denoted by capital
(lowercase) letters.

\subsection{Privileged gauge}

Using the previous harmonic expansions we can
decompose the $n$th-order gauge transformation
(\ref{highorderGI}) from the perturbations of the
metric in a generic gauge $\pert{n}{h}_{\mu\nu}$ to
the perturbations $\pert{n}{\cal K}_{\mu\nu}$ in some
privileged gauge, that we still have to specify:
\begin{eqnarray}\label{invariant1}
l \geq 0:\quad && \nonumber\\
\pert{n}{{\mathcal
K}}_l^m{}_{AB}&=&\pert{n}{H}_l^m{}_{AB}
+\pert{n}{P}_l^m{}_{A|B}+\pert{n}{P}_l^m{}_{B|A}\nonumber
\\
&+&\pert{n}{\cal H}_l^m{}_{AB},\\
\label{invariant4} \pert{n}{{\mathcal
K}}_l^m&=&\pert{n}{K}_l^m +2v^A\pert{n}{P}_l^m{}_A
-l(l+1)\pert{n}{P}_l^m\nonumber\\
&+&\pert{n}{\cal H}_l^m,\\
l \geq 1:\quad && \nonumber\\\label{invariant2}
\pert{n}{\cal K}_l^m{}_A&=&
\pert{n}{H}_l^m{}_{A}+\pert{n}{P}_l^m{}_{A}
+r^2\pert{n}{P}_l^m{}_{|A}\nonumber\\
&+&\pert{n}{\cal H}_l^m{}_A,\\\label{invariant3}
\pert{n}{\kappa}_l^m{}_{A}&=&
\pert{n}{h}_l^m{}_{A}+r^2\pert{n}{q}_l^m{}_{|A}
+\pert{n}{\check{h}}_l^m{}_A,\\
l \geq 2:\quad && \nonumber\\\label{invariant5}
\pert{n}{\tilde{\cal K}}_l^m&=&
\pert{n}{G}_l^m+2\pert{n}{P}_l^m
+\frac{1}{r^2}\pert{n}{\tilde{\cal H}}_l^m,\\
\label{invariant6}
\pert{n}{\kappa}_l^m&=&\pert{n}{h}_l^m+2r^2
\pert{n}{q}_l^m +\pert{n}{\check{h}}_l^m.
\end{eqnarray}
The cases $l=0,1$ are special inasmuch as some of the
tensor spherical harmonics vanish for these values of
$l$, and hence some of the relations that one obtains
for $l\geq 2$ are trivially void in these cases.

In general, the privileged gauge at order $n$ will be
determined by four conditions $\pert{n}{R}$, imposed
on the ten components of $\pert{n}{\cal K}_{\mu\nu}$.
These conditions must rigidly fix the gauge vector
(\ref{gaugeP}) as a function(al) of the generic metric
perturbation $\pert{n}{h}_{\mu\nu}$. In order to
obtain amenable gauge invariants, we would like to
select a gauge such that this functional dependence is
local with respect to the Lorentzian manifold ${\cal
M}^2$ (whereas the dependence on $S^2$ is determined
by means of the expansion in tensor harmonics). After
replacing the obtained gauge vector, the other six
components of $\pert{n}{\cal K}_{\mu\nu}$ will provide
then local (in ${\cal M}^2$) gauge invariants by
construction. For simplicity, we will employ from now
on the adjective ``local'' exclusively in the sense of
locality in ${\cal M}^2$, so that no integration over
this manifold is involved.

From the above expressions, it is not difficult to see
that there are essentially two procedures to end with
a vector $\flow{n}{p}^{\mu}$ whose components present
a polynomic dependence on those of $\pert{n}{\cal
K}_{\mu\nu}$ and $\pert{m}{h}_{\mu\nu}$, for $m\leq
n$. In both cases, the components $\pert{n}{P}_l^m$
and $\pert{n}{q}_l^m$ are found solving Eqs.
(\ref{invariant5}) and (\ref{invariant6}),
respectively. But for the vector component
$\pert{n}{P}_l^m{}_A$ one has two options. One can
fully determine it from Eq. (\ref{invariant2}), or one
can instead obtain the projection
$v^A\pert{n}{P}_l^m{}_A$ from Eq. (\ref{invariant4})
and the remaining projection $t^A\pert{n}{P}_l^m{}_A$
from Eq. (\ref{invariant2}), with $t^A$ being any
vector transverse to $v^A$.

This fact suggests two possible sets of conditions to
select our privileged gauge, both of them imposed on
the tensor $\pert{n}{\cal K}_{\mu\nu}$. The first one
consists of the following requirements
\begin{eqnarray}
l\ge 1: && \pert{n}{\cal K}_l^m{}_A=0, \\
l\ge 2: && \pert{n}{\tilde{\cal K}}_l^m=0,
\qquad\pert{n}{\kappa}_l^m=0 .
\end{eqnarray}
This is the RW gauge used at first order
\cite{ReWh57}. This gauge has been extensively
employed in the study of perturbations of spherical
backgrounds and leads to a full metric whose pullback
satisfies \cite{BMM06}
\begin{equation} \label{generalRW}
(\overline{\phi}_\varepsilon^*g_\varepsilon)_{Ab:c}\,
g^{bc}=0, \qquad
(\overline{\phi}_\varepsilon^*g_\varepsilon)_{ab} =
\tilde{K}_\varepsilon g_{ab},
\end{equation}
for some generic scalar field $\tilde{K}_\varepsilon$
defined in the background four-dimensional manifold
$\cal M$. It is well known that this gauge is not
completely rigid because it does not impose any
restriction on some of the $l=0,1$ components of the
metric perturbations, at any order. For the rest of
this subsection, we will restrict our analysis of
gauge invariance to transformations whose generators
do not contain the corresponding $l=0,1$ modes (both
at order $n$ and lower orders). We defer the analysis
of the general case to the next subsection.

With the choice of the RW gauge, Eqs.
(\ref{invariant2}), (\ref{invariant5}), and
(\ref{invariant6}) are the conditions that must be
solved to obtain the gauge vector $\flow{n}{p}^{\mu}$,
which turns out to be
\begin{eqnarray}\label{RWvectorn1}
\pert{n}{P}_l^m{}_A&=&\frac{r^2}{2}\left(\pert{n}{G}_l^m{}
+\frac{1}{r^2}\pert{n}{\tilde{\cal
H}}_l^m{}\right)_{|A}
-\pert{n}{H}_l^m{}_A\nonumber\\
&-&\pert{n}{\cal H}_l^m{}_A,\\\label{RWvectorn2}
\pert{n}{P}_l^m&=&-\frac{1}{2}\left(\pert{n}{G}_l^m+
\frac{1}{r^2}\pert{n}{\tilde{\cal
H}}_l^m\right),\\\label{RWvectorn3}
\pert{n}{q}_l^m&=&-\frac{1}{2r^2}\left(\pert{n}{h}_l^m
+\pert{n}{\check{h}}_l^m\right).
\end{eqnarray}
Substituting these expressions in Eqs.
(\ref{invariant1}), (\ref{invariant4}), and
(\ref{invariant3}) we obtain explicitly the high-order
metric gauge invariants.

The axial part of the second algebraic gauge is
equivalent to the RW gauge, and therefore
$\pert{n}{q}_l^m$ is given again by Eq.
(\ref{RWvectorn3}). As a consequence, the axial
invariant is the same as before, namely
$\pert{n}{\kappa}_l^m{}_A$. However, the polar part is
now determined by the following conditions:
\begin{eqnarray}\label{newgauge}
l\ge 0: && \pert{n}{\cal K}_l^m=0, \\
l\ge 1: && \pert{n}{\cal K}_l^m{}_A\, t^A=0, \\
l\ge 2: && \pert{n}{\tilde{\cal K}}_l^m=0,
\end{eqnarray}
which generalize the Campolattaro-Thorne gauge $K=0$
\cite{CaTh70} for the $l=1$ component of first-order
perturbations. The exact gauge would be
\begin{equation}
(\overline{\phi}_\varepsilon^*g_\varepsilon)_{Ab:c} \,
g^{bc}= \epsilon_{AB}t^B\tilde H_\varepsilon, \qquad
(\overline{\phi}_\varepsilon^*g_\varepsilon)_{ab} =
g_{ab},
\end{equation}
for some scalar field $\tilde H_\varepsilon$ in the
background $\cal M$. The above comments about the lack
of rigidity are also applicable to this gauge. That
is, the gauge generators are assumed to have no
component with harmonic label $l=0,1$.

For the polar components of the vector
$\flow{n}{p}^{\mu}$, we solve conditions
(\ref{newgauge}) and obtain
\begin{eqnarray}
v^AP_l^m{}_A&=&
-\frac{l(l+1)}{4}\left(\pert{n}{G}_l^m+
\frac{1}{r^2}\pert{n}{\tilde{\cal
H}}_l^m\right)\nonumber
\\&-&\frac{1}{2}\pert{n}{K}_l^m,\\
t^AP_l^m{}_A&=&\frac{r^2}{2}t^A\left(\pert{n}{G}_l^m+
\frac{1}{r^2}\pert{n}{\tilde{\cal H}}_l^m\right)_{|A}
\nonumber\\&-&t^A\left(\pert{n}{H}_l^m{}_A+
\pert{n}{\cal H}_l^m{}_A\right),\\
P_l^m&=&-\frac{1}{2}\left(\pert{n}{G}_l^m+
\frac{1}{r^2}\pert{n}{\tilde{\cal H}}_l^m\right).
\end{eqnarray}
Substituting the above relations in the respective
definitions (\ref{invariant1}) and (\ref{invariant2}),
one arrives at the corresponding polar invariants
${\cal K}_l^m{}_{AB}$ and $v^A{\cal K}_l^m{}_A$.

In conclusion, we have seen that there are two choices
of gauge that lead to a polynomic and local expression
of the gauge vector $\flow{n}{p}^\mu$ in terms of the
metric perturbation $\pert{n}{h}_{\mu\nu}$ in an
arbitrary gauge. For both of them, one obtains a set
of local gauge invariants, under the restriction $l\ge
2$ on the modes of the gauge generators, at all
perturbative orders. These two sets are inequivalent.
Nevertheless, the RW gauge does not need the
introduction by hand of a vector field $t_A$
transverse to $v_A$, so that one can consider it to be
more natural, at least as far as this kind of
geometric considerations are concerned. Because of
these reasons, we choose the RW gauge as our
privileged gauge from now on. In Paper I we already
computed the evolution equations for the perturbations
in the RW gauge. Of course, those equations are also
valid for the gauge invariants that are tied to this
gauge.

The fact that there exist privileged gauges that
provide local gauge invariants is related to the
spherical symmetry of the background, which allows the
choice of an adapted system of coordinates in which
the metric has a block-diagonal form. Besides, Kodama,
Ishibashi, and Seto \cite{KIS} have shown that it is
possible to generalize the first-order GS invariants
for any metric which can be written in block-diagonal
form in an $(m+n)$-dimensional manifold ${\cal
M}^m\times {\cal K}^n$ with a maximally symmetric
block in ${\cal K}^n$. This means that a
higher-dimensional form of the RW gauge is always
attainable, though the corresponding conditions on the
metric perturbations may be different.

\subsection{The particular cases $l=0,1$}

It is well-known that, at first order, one cannot
construct local gauge invariants for $l=0,1$ by the
methods explained in the previous subsection. This is
because some of the equations
(\ref{invariant1}--\ref{invariant6}) are not present
in those cases, and therefore it is not possible to
attain a local expression for the components of the
gauge vector $\pert{n}{p}^{\mu}$ in terms of those of
the metric perturbations. Of course, gauge conditions
different from the RW ones can be imposed on the
metric perturbations and hence one can obtain from
them the associated gauge invariants, but these will
be nonlocal because the gauge vector will be given by
an integral expression (over ${\cal M}^2$) of the
metric perturbations. Whether this is useful or not
will depend on the particular application that one is
studying.

The same obstruction appears as well at every order in
perturbation theory, so that one cannot get local
gauge invariants for $l=0,1$ at any order. However,
mode coupling makes the problem worse: the existence
of lower-order modes with $l=0,1$ may prevent the
construction of higher-order local gauge invariants
with $l\ge 2$. This will happen when such lower-order
modes have a nonzero contribution to the sources
${\cal H}$ of Eqs.
(\ref{invariant1}--\ref{invariant6}).

The restriction on the gauge generators introduced in
the previous subsection is a valid alternative for
first-order perturbation theory: the GS gauge
invariants remain unchanged under the restricted group
$\pert{1}{\cal G}^{\diamond}$, where the diamond
indicates that no first-order generator with $l=0,1$
is included. However, removing all $l=0,1$ generators
is already inconsistent at second order if we demand
invariance under a group of transformations, because
the $l=0,1$ components of the vectors $\flow{2}{\xi}$
will be unavoidably generated by coupling of
first-order gauge modes [cf. composition
(\ref{coupling1})]. Fortunately, those offending gauge
modes act only on the $l=0,1$ second-order
perturbations, for which invariants cannot be
constructed anyway. All other second-order
perturbations admit a gauge-invariant form, as given
in the previous subsection, under the gauge group
$\pert{2}{\cal G}^{\diamond}$ where again the diamond
denotes that no first-order $l=0,1$ gauge mode is
included, but all other first and second-order
(including $l=0,1$) modes are allowed. That is one of
the main results of this investigation:
gauge-invariant perturbations can be consistently and
simultaneously constructed for all $l\ge 2$ modes at
second order, all being invariant under the group
$\pert{2}{\cal G}^{\diamond}$.

The situation at third and higher orders is more
restrictive. As a summary, working at order $n>2$ only
a finite number of gauge generators can be included in
the invariance group at orders $1,...,n-2$. The
unavoidable presence of second-order gauge modes with
$l=0,1$ couples to any metric perturbation at order
$m$ and with harmonic label $l$, preventing the
construction of a gauge invariant of order $m+2$ and
label $l$ or $l\pm 1$, with respect to those gauge
modes. The detailed analysis is shown in Appendix
\ref{coupling}.

\subsection{First-order gauge invariants}

We now particularize our discussion to first-order
perturbations and reproduce the GS first-order gauge
invariants \cite{GeSe79}, which are a generalization
of the invariants introduced by Moncrief \cite{Mon73}
for a Schwarzschild background.

Since the source $\pert{1}{\cal H}_{\mu\nu}$ vanishes,
from Eqs. (\ref{invariant1}), (\ref{invariant4}), and
(\ref{invariant3}) one gets the metric invariants
\begin{eqnarray}
\pert{1}{{\mathcal K}}_l^m{}_{AB}&=&
\pert{1}{H}_l^m{}_{AB}
+\pert{1}{P}_l^m{}_{A|B}+\pert{1}{P}_l^m{}_{B|A},\\
\pert{1}{{\mathcal K}}_l^m&=&\pert{1}{K}_l^m+
2v^A\pert{1}{P}_l^m{}_A\nonumber\\
&-&l(l+1)\pert{1}{P}_l^m,\\
\pert{1}{\kappa}_l^m{}_{A}&=&\pert{1}{h}_l^m{}_{A}
+r^2\pert{1}{q}_l^m{}_{|A},
\end{eqnarray}
where
\begin{eqnarray}
\pert{1}{P}_l^m{}_A&=&\frac{r^2}{2}\pert{1}{G}_l^m{}_{|A},\\
\pert{1}{P}_l^m&=&-\frac{1}{2}\pert{1}{G}_l^m,\\
\pert{1}{q}_l^m&=&-\frac{1}{2r^2}\pert{1}{h}_l^m.
\end{eqnarray}
Using these expressions for the gauge vector, it is
possible to construct the first-order gauge invariants
corresponding to any other (matter or metric)
tensorial object.

\subsection{Second-order gauge invariants}

The second-order metric invariants can be obtained by
substituting expressions
(\ref{RWvectorn1}--\ref{RWvectorn3}) in Eqs.
(\ref{invariant1}), (\ref{invariant4}), and
(\ref{invariant3}) for $n=2$, whereas the matter
invariants, constructed from the perturbations
$\pert{n}{\Psi}_{\mu\nu}\equiv \Delta^n[t_{\mu\nu}]$,
are encoded in the tensor $\pert{2}{T}_{\mu\nu}$
defined by
\begin{eqnarray}
\pert{2}{T}_{\mu\nu}&=&\pert{2}{\Psi}_{\mu\nu}
+{\mathcal L}_{\pert{2}{p}}\Psi_{\mu\nu} +{\mathcal
L}^2_{\pert{1}{p}}\Psi_{\mu\nu}
\nonumber\\
&+&2{\mathcal L}_{\pert{1}{p}}\pert{1}{\Psi}_{\mu\nu}.
\end{eqnarray}
For the sake of brevity we will only provide
explicitly the form of the metric gauge invariants,
but not that of the matter invariants, which can be
computed along the same lines.

Our aim is to compute the harmonic coefficients
(\ref{gaugedec}) of the source
\begin{equation}\label{source2}
\pert{2}{\cal H}_{\mu\nu}\equiv{\mathcal
L}^2_{\pert{1}{p}}g_{\mu\nu} +2{\mathcal
L}_{\pert{1}{p}}\pert{1}{h}_{\mu\nu}.
\end{equation}
These coefficients will have the general form (we
obviate the $n=2$ label for simplicity):
\begin{eqnarray}\label{calHsource1}
{\cal H}_l^m{}_{AB}&=&\sum_{\bar{l},\hat{l}}
\sum_{\bar{m},\hat{m}}
\calH{\epsilon}{\bar{l}}{\bar{m}}{\hat{l}}
{\hat{m}}{l}{m}{}_{AB},\\\label{calHsource2} {\cal
H}_l^m{}_{A}&=&\sum_{\bar{l},\hat{l}}
\sum_{\bar{m},\hat{m}}
\calH{\epsilon}{\bar{l}}{\bar{m}}{\hat{l}}
{\hat{m}}{l}{m}{}_{A},\\\label{calHsource3} {\cal
H}_l^m&=&\sum_{\bar{l},\hat{l}} \sum_{\bar{m},\hat{m}}
\calH{\epsilon}{\bar{l}}{\bar{m}}{\hat{l}}
{\hat{m}}{l}{m},\\\label{calHsource4} \tilde{\cal
H}_l^m&=&\sum_{\bar{l},\hat{l}} \sum_{\bar{m},\hat{m}}
\calHtilde{\epsilon}{\bar{l}}{\bar{m}}{\hat{l}}
{\hat{m}}{l}{m},\\\label{calHsource5}
\check{h}_l^m{}_A&=&-i\sum_{\bar{l},\hat{l}}
\sum_{\bar{m},\hat{m}}
\calH{-\epsilon}{\bar{l}}{\bar{m}}{\hat{l}}
{\hat{m}}{l}{m}{}_{A},\\\label{calHsource6}
\check{h}_l^m&=&-i\sum_{\bar{l},\hat{l}}
\sum_{\bar{m},\hat{m}}
\calH{-\epsilon}{\bar{l}}{\bar{m}}
{\hat{l}}{\hat{m}}{l}{m},
\end{eqnarray}
with $\bar{l}$ and $\hat{l}$ being the harmonic labels
corresponding to the two first-order modes that get
coupled. They are independent and run over all
nonnegative integers with the restriction $|\bar
l-\hat l|\leq l \leq |\bar l+\hat l|$. In addition,
one must impose the usual restrictions on the values
of the labels $\bar{m}$ and $\hat{m}$. We have also
defined the alternating sign $\epsilon\equiv(-1)^{\bar
l+\hat l-l}$. As it is shown in Appendix
\ref{coupling}, this alternating sign encodes the
mixing of polarities caused by the product of tensor
harmonics in higher-order perturbation theory. For a
detailed explanation of this effect and the use of the
sign $\epsilon$ we refer the reader to Paper I. As in
that reference, all computations have been carried out
with the computer algebra system {\em xTensor}
\cite{xTensor}.

From now on and without loss of generality, we will
suppose that there are only two first-order modes that
will be denoted as $(\hat {\cal K}, \hat p)$ and
$(\overline {\cal K}, \overline p)$. The tensorial
sources are given by
\begin{widetext}
\begin{eqnarray}\label{calHABreal}
\calH{+}{\bar{l}}{\bar{m}}
{\hat{l}}{\hat{m}}{l}{m}{}_{AB}&=& -4 \E{1}{{\hat
l}}{{\hat m}}{-1}{{\bar l}}{{\bar m}}{l} \left\{ {\bar
P}{\hat{\cal K}}_{AB}+2{\bar
q}_{|(B}{\hat\kappa}_{A)}- ({\bar P}{\hat
P}_{(B})_{|A)} -r^2{\hat P}_{|A}{\bar P}_{|B}-r^2{\hat
q}_{|A}{\bar q}_{|B} \right\}
\nonumber\\
&+& 2 \E{0}{{\hat l}}{{\hat m}}{0}{{\bar l}}{{\bar
m}}{l} \left\{ 2{\bar P}^C{}_{|(A}{\hat{\cal
K}}_{B)C}+{\bar P}^C{\hat{\cal K}}_{AB|C} -{\hat
P}^C{}_{|(B}{\bar P}_{A)|C}-{\hat P}^C{}_{|A} {\bar
P}_{C|B}-{\hat P}^C{\bar P}_{(A|B)C} \right\},
\\\label{calHABimag}
\calH{-}{\bar{l}}{\bar{m}}{\hat{l}}
{\hat{m}}{l}{m}{}_{AB}&=& -4i \E{1}{{\hat l}}{{\hat
m}}{-1}{{\bar l}}{{\bar m}}{l} \left\{{\hat
q}{\bar{\cal K}}_{AB}+2{\bar P}_{|(A}{\hat\kappa}_{B)}
-2r^2{\hat q}_{|(A}{\bar P}_{|B)} +{\bar q}{\hat
P}_{(A|B)} +{\bar q}_{|(A}{\hat P}_{B)} \right\}.
\end{eqnarray}
\end{widetext}
The real coefficients $E$ come from the product of
harmonics \cite{BMM06} that is implicit in Eq.
(\ref{source2}). They are defined as
\begin{equation}\label{coeffE}
\E{s'}{l'}{m'}{s}{l}{m}{l''} \equiv
\frac{k(l',|s'|)k(l,|s|)}{k(l'',|s+s'|)}
\C{l'}{m'}{l}{m}{l''}{m'+m}
\C{l'}{s'}{l}{s}{l''}{s'+s},
\end{equation}
where $\C{l_1}{m_1}{l_2}{m_2}{l}{m_1+m_2}$ are
Clebsch-Gordan coefficients and we have defined the
normalization factor
\begin{equation}
k(l,s) \equiv \sqrt{\frac{(2l+1)(l+s)!}{ \,\,2^{s+2}\,
\pi\,(l-s)!}}.
\end{equation}

The vectorial sources are decomposed as
\begin{widetext}
\begin{eqnarray}\label{calHAbreal}
\calH{+}{\bar{l}}{\bar{m}}{\hat{l}}{\hat{m}}{l}{m}{}_{A}&=&
\E{-1}{{\hat l}}{{\hat m}}{2}{{\bar l}}{{\bar m}}{l}
\left\{ -2({\bar q}{\hat\kappa}_A+{\hat
q}{\bar\kappa}_A) +({\hat P}_A{\bar P}+{\bar P}_A{\hat
P}) +r^2(3{\bar P}{\hat P}_{|A}+{\hat P}{\bar P}_{|A})
+ 2{\bar q}(r^2{\hat q})_{|A} +r^6\left(\frac{{\hat
q}{\bar q}}{r^4}\right)_{|A} \right\} \nonumber\\ &+&
\frac{1}{2}\E{1}{{\hat l}}{{\hat m}}{0}{{\bar
l}}{{\bar m}}{l} \bigg\{ 2{\bar l}({\bar l}+1)({\hat
q}{\bar\kappa}_A-{\bar q}{\hat\kappa}_A) +4{\hat
P}^B{\bar{\cal K}}_{AB}+4r^2{\hat P}_{|A}{\bar{\cal
K}} \nonumber\\ &+&\left. {\bar l}({\bar
l}+1)\left({\bar P}{\hat P}_A+{\hat P}{\bar P}_A
+3r^2{\bar P}{\hat P}_{|A}+r^2{\hat P}{\bar P}_{|A}
+r^2{\bar q}{\hat q}_{|A}-r^2{\hat q}{\bar q}_{|A}
\right) \right.\nonumber\\ &-& 2r^2\left[\left({\bar
P}^B{\hat P}_{|B}\right)_{|A} +4v^B{\bar P}_B{\hat
P}_{|A}\right] -2\left( {\bar P}^B{\hat
P}_{A|B}+2{\hat P}^B{\bar P}_{B|A} +{\hat P}^B{\bar
P}_{A|B} \right) \bigg\} ,\\\label{calHAbimag}
\calH{-}{\bar{l}}{\bar{m}}{\hat{l}}
{\hat{m}}{l}{m}{}_{A}&=& -i\E{-1}{{\hat l}}{{\hat
m}}{2}{{\bar l}}{{\bar m}}{l} \left\{ 2\left({\hat
P}{\bar\kappa}_A-{\bar P}{\hat\kappa}_A\right) +{\hat
q}{\bar P}_A-{\bar q}{\hat P}_A +r^2{\hat q}{\bar
P}_{|A}+3r^2{\bar P}{\hat q}_{|A} - 3r^2{\bar q}{\hat
P}_{|A}-r^2{\hat P}{\bar q}_{|A}
\right\}\nonumber\\
&+&i\E{0}{{\hat l}}{{\hat m}}{1}{{\bar l}}{{\bar
m}}{l} \left\{ 2r^2{\hat{\cal K}}{\bar q}_{|A} -{\hat
l}({\hat l}+1)\left( {\hat P}{\bar\kappa}_A+{\bar
P}{\hat\kappa}_A \right) +2\left( {\bar\kappa}^B{\hat
P}_{B|A}+{\hat
P}^B{\bar\kappa}_{A|B} \right) \right.\nonumber\\
&+&\left. \frac{{\hat l}({\hat l}+1)}{2}\left(
r^2{\bar q}{\hat P}_{|A}-r^2{\hat q}{\bar P}_{|A}
+3r^2{\hat P}{\bar q}_{|A}+r^2{\bar P}{\hat q}_{|A}
-{\hat q}{\bar
P}_A+{\bar q}{\hat P}_A \right) \right.\nonumber\\
&-&\left.r^2\left( {\hat P}^B{\bar q}_{|BA} +{\hat
P}^B{}_{|A}{\bar q}_{|B} +4v^B{\hat P}_B{\bar q}_{|A}
\right) \right\},
\end{eqnarray}
and finally the four scalar sources are given by
\begin{eqnarray}\label{calHabgreal}
\calH{+}{\bar{l}}{\bar{m}}{\hat{l}}{\hat{m}}{l}{m}&=&
-4\E{2}{{\hat l}}{{\hat m}}{-2}{{\bar l}}{{\bar m}}{l}
\left( {\hat q}{\bar q}+{\hat P}{\bar P}
\right)+\E{1}{{\hat l}}{{\hat m}}{-1}{{\bar l}}{{\bar
m}}{l} \left\{ -4{\hat P}{\bar{\cal K}}-\left[{\hat
l}({\hat l}+1)+{\bar l}({\bar l}+1) \right]{\hat
P}{\bar P}+\frac{2}{r^2} {\hat P}^A(r^2{\bar
P})_{|A}+\frac{2}{r^2}{\hat P}^A {\bar
P}_A\right\}\nonumber\\\nonumber &+&\E{0}{{\hat
l}}{{\hat m}}{0}{{\bar l}}{{\bar m}}{l} \bigg\{
-2{\bar l}({\bar l}+1){\bar P}{\hat{\cal
K}}+\frac{2}{r^2} {\bar P}^A(r^2{\hat{\cal
K}})_{|A}+\frac{{\hat l}({\hat l}+1)}{r^4} {\bar
P}^A(r^4{\hat P})_{|A}\\
&-&{\hat l}({\hat l}+1){\bar l}({\bar l}+1){\hat
P}{\bar P} -2{\hat P}^A\left[ ({\bar
P}^Bv_B)_{|A}+2{\bar P}^Bv_Av_B \right]
\bigg\},\\\label{calHabgimag}
\calH{-}{\bar{l}}{\bar{m}}{\hat{l}}{\hat{m}}{l}{m}&=&
8i\E{2}{{\hat l}}{{\hat m}}{-2}{{\bar l}}{{\bar
m}}{l}{\hat P}{\bar q} -2i\E{1}{{\hat l}}{{\hat
m}}{-1}{{\bar l}}{{\bar m}}{l} \left\{
\frac{2}{r^2}{\bar P}^A{\hat\kappa}_A-2{\bar
q}{\hat{\cal K}}- {\hat l}({\hat l}+1){\hat P}{\bar q}
+\frac{1}{r^2}{\hat P}^A(r^2{\bar q})_{|A} \right\}
,\\\label{calHabZreal}
\calHtilde{+}{\bar{l}}{\bar{m}}{\hat{l}}
{\hat{m}}{l}{m}&=&2r^2\E{3}{{\hat l}}{{\hat m}}{-1}
{{\bar l}}{{\bar m}}{l} \left( {\hat q}{\bar q}+{\hat
P}{\bar P} \right)+ r^2\E{1}{{\hat l}}{{\hat
m}}{1}{{\bar l}}{{\bar m}}{l} \left\{ ({\hat
l}+2)({\hat l}-1)\left[{\hat P}{\bar P}-{\hat q}{\bar
q}\right] -2{\hat P}^A{\bar P}_{|A}-\frac{2}{r^2}{\hat
P}^A{\bar P}_A \right\} \nonumber\\ &+&
2r^2\E{2}{{\hat l}}{{\hat m}}{0}{{\bar l}}{{\bar
m}}{l} \left\{ 2{\hat P}{\bar{\cal K}}+{\bar l}({\bar
l}+1)\left[ {\hat q}{\bar q}+2{\hat P}{\bar P} \right]
-\frac{1}{r^4}{\bar P}^A\left(r^4{\hat P}\right)_{|A}
\right\},
\\\label{calHabZimag}
\calHtilde{-}{\bar{l}}{\bar{m}}{\hat{l}}
{\hat{m}}{l}{m}&=&2ir^2\E{3}{{\hat l}}{{\hat
m}}{-1}{{\bar l}} {{\bar m}}{l}\left( {\hat q}{\bar
P}-{\hat P}{\bar q} \right) +i\E{2}{{\hat l}}{{\hat
m}}{0}{{\bar l}}{{\bar m}}{l} \left\{ 4r^2{\hat
q}{\bar{\cal K}}+2r^2{\bar l}({\bar l}+1)\left[ 2{\bar
P}{\hat q}-{\hat P}{\bar q} \right]
-\frac{2}{r^2}{\bar P}^A (r^4{\hat q})_{|A} \right\}\nonumber\\
&+&i\E{1}{{\hat l}}{{\hat m}}{1}{{\bar l}}{{\bar
m}}{l} \left\{ 4{\bar P}^A{\hat\kappa}_A+({\hat
l}+2)({\hat l}-1)r^2\left[ {\hat P}{\bar q}+{\bar
P}{\hat q} \right]-2r^2{\hat P}^A{\bar q}_{|A}
\right\}.
\end{eqnarray}
\end{widetext}
These formulas allow us to construct the
gauge-invariant form (under the restricted group
$\pert{2}{\cal G}^{\diamond}$ of transformations) of
any second-order metric perturbation in a covariant
way. Note that we have not restricted the harmonic
labels corresponding to the first-order metric
perturbations. In particular we allow the values $l=0$
and $l=1$ for them, although then some of the harmonic
coefficients in the formulas will vanish.

Appendix \ref{GaPr} studies the particularization of
these gauge invariants to the case of second-order
perturbations of a vacuum in Schwarzschild coordinates
when one has at first order a single polar mode with
$l=2$, comparing the results with those of Ref.
\cite{GaPr00}.

\section{Conclusions}
\label{conclusions}

We have completed the construction of a framework for
the analysis of high-order gauge-invariant
nonspherical perturbations of a general spherical
spacetime, with any type of matter content. Our work
started in Paper I by generalizing to higher orders
the GS formalism for first-order perturbations,
although the perturbations were described in that work
in a particular gauge. In this article we have given
explicit algorithms to construct gauge-invariant
perturbations at any order, focusing again on the
second order, for which we have provided full
expressions of the metric gauge invariants.

Before doing that, it has been necessary to discuss in
detail what we mean by gauge invariance. The simplest
or most natural form of gauge invariance (introduced
by Sachs \cite{Sac64}) turns out to be too restrictive
in spherical symmetry, even more when dealing with
high-order perturbations. We have used an alternative
form of gauge invariance, based on the work by
Moncrief \cite{Mon73} on perturbations about
Schwarzschild. Borrowing ideas from Bardeen
\cite{Bar80}, we interpret Moncrief's gauge invariants
as being associated with a choice of preferred gauge.
This reinterpretation has allowed us to extend the
construction of gauge invariants to higher orders in
perturbation theory, our main goal in this work.

The question then arises of what has been really
gained in defining gauge invariants if these are
associated with a preferred gauge. First of all, it is
important to emphasize that these are true invariants,
in the sense that they do not change under gauge
transformations of the perturbations, and therefore in
a given problem we do not have to worry about the
gauge in which the perturbations actually are
described. Secondly, the freedom to choose the
preferred gauge is severely restricted in practice by
the requirement of obtaining amenable explicit
expressions for the invariants. For example in
spherical symmetry the demand of arriving at explicit
expressions that are local with respect to the
dependence on ${\cal M}^2$ almost singles out the RW
gauge, which essentially renders the associated
invariants unique. For nonsymmetric background
spacetimes this type of gauge invariants, with an
explicit and manageable form in terms of the
perturbations in a generic gauge, will generically not
exist.

Unfortunately the RW gauge cannot be imposed on
perturbations with $l=0$ or $l=1$ at all orders, and
this obstacle propagates to higher values of $l$
through mode coupling. In this way, the invariants
associated with the RW gauge remain unaltered under
the set of gauge transformations whose generators
exclude all the modes with harmonic label $l$ equal to
0 or 1. In general, however, this set of
transformations does not have group structure owing to
mode coupling. Nevertheless, at second order, the
group $\pert{2}{\cal G}^{\diamond}$ (which does not
contain the $l=0,1$ first-order gauge generators)
allows the construction of all gauge invariants with
$l\ge 2$ both at first and second orders, invariants
that are exclusively associated with the RW gauge and
whose expression we have been able to deduce for the
most general case (i.e., for any matter model and any
background spacetime).

Our framework is now ready to be applied to specific
matter models and we are currently working on its
application to the study of second-order perturbations
of a time-dependent ball of perfect fluid, with a
general (2-parameter) equation of state.

\acknowledgments

D.B. acknowledges financial support from the FPI
program of the Regional Government of Madrid.
J.M.M.-G. acknowledges financial aid provided by the
I3P framework of CSIC and the European Social Fund.
This work was supported by the Spanish MEC Project
FIS2005-05736-C03-02.

\appendix

\section{Mode coupling among gauge generators}\label{coupling}

We have constructed second-order gauge invariants
under the group $\pert{2}{\cal G}^{\diamond}$ starting
just from the non(fully-)rigid RW gauge. This is not
possible at higher orders and actually one has to
restrict to a finite set of lower-order modes both in
the gauge generators and in the perturbations in order
to define some form of gauge invariance. This is
because the presence at first order of any gauge mode
$l\geq 2$ will generate, just by self-coupling, the
second-order modes with harmonic labels 0 and 1. It
will then be imposible to construct the
gauge-invariant form of a third-order perturbation
whose source ${\cal H}$ contains a term coupling any
of these second-order $l=0,1$ modes with any
first-order mode. But those sources generically
contain all possible couplings, and so only a problem
in which we restrict the number of first-order gauge
modes allows some form of gauge-invariance at third
order. In this appendix we will analyze generic mode
coupling around spherical symmetry, starting at second
order and then proceeding to higher orders. We will
later give some bounds on the number of modes that can
be present at first order to allow for the
construction of a $n$th-order mode with label $l$.

The second-order $l$-mode will get a contribution from
a pair of first-order modes $\hat l$ and $\bar l$ if
two conditions are obeyed. On the one hand, the
harmonic labels must be related by the usual
composition formula
\begin{equation}\label{composition}
|\hat l-\bar l|\leq l \leq \hat l+\bar l.
\end{equation}
On the other hand, mode coupling must conserve parity.
To any harmonic coefficient with label $l$, we
associate a polarity sign $\sigma$ such that, under
parity, the harmonic changes by a sign $\sigma(-1)^l$.
Polar (axial) harmonics have $\sigma=+1$
($\sigma=-1$). Then, parity conservation implies the
second condition:
\begin{equation} \label{parity}
(-1)^{\bar l+\hat l-l}\equiv \epsilon
=\sigma\bar\sigma\hat\sigma,
\end{equation}
where we have made use of the alternating sign
$\epsilon$. There is a special case in which the
coupling of two modes satisfying Eqs.
(\ref{composition}) and (\ref{parity}) does not
contribute to a second-order mode, and the reason
comes from the properties of the Clebsch-Gordan
coefficients that appear in the product formula for
the tensor harmonics (see Paper I). In axisymmetry
($\bar m=\hat m=0$) the Clebsch-Gordan coefficients,
and as a consequence the E-coefficients that couple
the modes [defined in Eq. (\ref{coeffE})], vanish if
$\bar l+\hat l+l$ is odd. A highly geometric
derivation of the perturbative structure under the
assumption of axisymmetry is given in Ref.
\cite{Car00}.

This analysis can be extended to higher orders. In
particular, the parity condition will be that a
collection of $k$ modes with harmonic labels
$\{l_1,...,l_k\}$ and polarities
$\{\sigma_1,...,\sigma_k\}$ will contribute to the
mode $(l,\sigma)$ only if $(-1)^l\sigma =
\Pi_{i=1}^{k}(-1)^{l_i}\sigma_i$.

Let us finally consider the case in which we have a
first-order finite collection of modes with their
harmonic labels taking all the values from $l=2$ to
$l= l_{\rm max}$, with contributions from both the
polar and the axial sectors. Coupling of these modes
at order $n$ will generate some new modes, following
the above rules, so that the highest value of their
harmonic label will be $n l_{\rm max}$. The
construction of the gauge invariants under the
corresponding group of transformations and tied to the
RW gauge is only guaranteed for those modes with
harmonic label greater than $(n-2) l_{\rm max} + 1$.
This number comes from the coupling of the
$(n-2)$th-order $(n-2) l_{\rm max}$-mode with the
second-order $l=1$ mode.

\section{Comparison with the results of
Garat and Price}\label{GaPr}

Garat and Price calculated the contribution of two
axisymmetric ($\bar m=\hat m=0$) first-order polar
quadrupoles ($\bar l=\hat l=2$) to the $l=2$ multipole
of the second-order Zerilli gauge invariant
\cite{GaPr00}. Their calculation was done in
Schwarzschild $(t, r)$ coordinates. In order to
translate their coordinate-dependent expressions to
our covariant notation, we introduce the following
frame of vectors:
\begin{equation}
r^A\equiv f \left(\frac{\partial}{\partial
r}\right)^A, \qquad t^A\equiv f^{-1}
\left(\frac{\partial}{\partial t} \right)^A,
\end{equation}
where $f\equiv\sqrt{r^{|A}r_{|A}}=\sqrt{1-2M/r}$ and
$M$ is the background black-hole mass. They are
normalized in the following way,
\begin{equation}
r^Ar_A=1,\qquad t^At_A=-1,\qquad r^At_A=0.
\end{equation}
The metric and Levi-Civita tensor are given by
\begin{eqnarray}
g_{AB}&=&-t_At_B+r_Ar_B,\\
\epsilon_{AB}&=&r_At_B-r_Bt_A.
\end{eqnarray}
Note that the relation between our frame vectors and
the coordinate 1-forms is
\begin{equation}
r_A=f^{-1}r_{|A},\qquad t_A=-ft_{|A}.
\end{equation}

Another question to consider in order to compare our
results with those of Garat and Price is the
definition of the tensor spherical harmonics. Firstly,
their normalization is different, which makes their
harmonic coefficients to be a factor
$\alpha^{-1}\equiv 2\sqrt{\pi/5}$ greater than ours.
Secondly, there is a mix of harmonic coefficients $G$
and $K$ because Garat and Price employed a harmonic
proportional to ${Y_2^0}_{:ab}$ instead of using
$Z_2^0{}_{ab}$.

The translation from the harmonic coefficients used by
Garat and Price (which appear below in the left-hand
side and in their notation) to our coefficients is
\begin{eqnarray}
H_0^{(n)} &\rightarrow & \alpha
t^At^B\pert{n}{H}_{AB},\\
H_1^{(n)} &\rightarrow & \alpha
r^At^B\pert{n}{H}_{AB},\\
H_2^{(n)} &\rightarrow & \alpha
r^Ar^B\pert{n}{H}_{AB},\\
h_0^{(n)} &\rightarrow & \alpha
f t^A\pert{n}{H}_A,\\
h_1^{(n)} &\rightarrow & \alpha
f^{-1} r^A\pert{n}{H}_A,\\
K^{(n)} &\rightarrow & \alpha
\left(3\pert{n}{G}+\pert{n}{K}\right),\\
G^{(n)} &\rightarrow & \alpha \pert{n}{G}.
\end{eqnarray}

In our notation, the $n$-th order Zerilli variable, in
terms of the GS invariants, is given by
\begin{eqnarray}\label{Zerilli}
\pert{n}{\Pi}_{{\rm
Z}}&\equiv&\frac{fr^2}{3M+\lambda}(fr^B\pert{n}
{\mathcal K}_{AB} -r\pert{n}{\mathcal
K}_{|A})r^A\nonumber\\&+&r\pert{n}{\mathcal K}.
\end{eqnarray}
In particular, at second order we could add to this
function any quadratic combination of first-order
modes, which would simply change the source term of
the Zerilli equation. Nevertheless, this is the
simplest Zerilli variable one can define. Using the
sources (\ref{calHABreal}--\ref{calHABimag}) and
(\ref{calHAbreal}--\ref{calHabZimag}) we have computed
the second-order GS invariants, compared our Zerilli
function (\ref{Zerilli}) with that found by Garat and
Price, and verified their formula (41) of Ref.
\cite{GaPr00}.

\end{document}